\documentclass[twocolumn,superscriptaddress,showpacs,preprintnumbers,amsmath,amssymb,aps,prb,floatfix,10pt]{revtex4-1}

\usepackage{graphicx}
\usepackage{color}
\usepackage{dcolumn} 
\usepackage{bm} 
\usepackage{epsfig}
\usepackage{multirow}

\newcommand{\comment}[1]{}

\begin{document}

\title{First-Principles Study of Electronic and Vibrational Properties of BaHfN$_2$}
\author{Amandeep Kaur}
\affiliation{Department of Physics, University of California, Davis, California 95616, USA}

\author{Erik R. Ylvisaker}
\affiliation{Department of Physics, University of California, Davis, California 95616, USA}

\author{Yan Li}
\affiliation{Department of Chemistry, University of California, Davis, California 95616, USA}

\author{Giulia Galli}
\affiliation{Department of Physics, University of California, Davis, California 95616, USA}
\affiliation{Department of Chemistry, University of California, Davis, California 95616, USA}

\author{Warren E. Pickett}
\affiliation{Department of Physics, University of California, Davis, California 95616, USA}

\begin{abstract}

The transition metal nitride BaHfN$_2$, which consists of weakly
bonded neutral slabs of closed shell ions, has structural and chemical
similarities to other layered nitrides which have impressive
superconducting T$_c$ when electron doped: A$_x$HfNCl, A$_x$ZrNCl,
A$_x$TiNCl, with $T_c= 25.5$, $15.2$ and $16.5$ K, respectively for
appropriate donor (A) concentrations $x$. These similarities suggest
the possibility of BaHfN$_2$ being another relatively high T$_c$
nitride upon doping, with effects of structure and the role of
specific transition metal ions yet to be understood.  We report
first-principles electronic structure calculations for stoichiometric
BaHfN$_2$ using density functional theory with plane-wave basis sets
and separable dual-space Gaussian pseudopotentials.  An indirect band
gap of 0.8 eV was obtained and the lowest conduction band is primarily
of Hf 5$d_{xy}$ character, similar to $\beta$-ZrNCl and
$\alpha$-TiNCl.  The two N sites, one in the Hf layer and another one
in the Ba layer, were found to have very anisotropic Born effective
charges (BEC):deviations from the formal charge (-3) are opposite for the
two sites, and opposite for the two orientations (in-plane, out of
plane).  LO-TO splittings and comparison of BECs and dielectric
constant tensors to those of related compounds are discussed, and the 
effect of electron doping on the zone-center phonons is reported.

\end{abstract}
\date{\today}
\maketitle

\section{Introduction}

High temperature superconductivity has been a puzzle since the
quasi-two-dimensional, doped insulating copper oxides were reported to
become superconducting with very high T$_c$'s.  Since then several
other layered transition metal oxides have been found to be good
superconductors although at relatively low temperature, for example
Li$_x$NbO$_2$\cite{GESE1990} and Na$_x$CoO$_2$\cite{TAKA2003} at about
5 K. The undoped parent compounds of the cuprate high-temperature
superconductors are magnetic insulators and their transition from a
magnetic insulator to a metal upon doping completely modifies their
electronic structure\cite{KAST1998}. These transition metal oxides
still attract a great deal of interest because the superconductivity
is not yet well understood.

Recently, interest has been growing for another class of layered
superconductors, the transition metal nitrides\cite{YAMA1996,GREG1998}
such as MNX (M=Ti, Zr, Hf; X=Cl, Br, I) and ternary transition metal
dinitrides AMN$_2$ (A=alkaline earth metal, M=Ti, Zr, Hf), some of
which have been reported to become superconducting with high T$_c$
values.  Superconductivity up to 12 K was first measured in
$\beta$-ZrNCl by Yamanaka {\it et al.} in 1996\cite{YAMA1996}, and
since then the highest T$_c$'s that have been measured for these
transition metal nitrides are as follows: $25.5$ K for
Li$_{0.48}$(THF)$_y$HfNCl\cite{YAMA1996}, $15.2$ K for intercalated
$\beta$-ZrNCl\cite{TAGU2006} and $16.5$ K recently
reported\cite{YAMA2009} for $\alpha$-TiNCl upon doping with Li.  These
electron-doped transition metal nitrides form a new and seemingly
unconventional class of high T$_c$ superconductors because, unlike the
transition metal oxides, the parent compounds are not Mott insulators.
The parent compounds for these layered quasi-2D nitrides are
non-magnetic ionic band insulators with a gap in the range of 2-4
eV.\cite{OHAS1989,FELS1999,HASE1999}

In these transition metal nitrides, the superconducting mechanism
presents a real conundrum.  Experimental measurements of the isotope
effect\cite{TAGU2007} in Li$_x$ZrNCl show a very weak dependence on
the N mass, suggesting that electron-phonon mediated pairing cannot
adequately account for the superconductivity in the MNX family.
Specific heat measurements on Li$_x$ZrNCl\cite{TAGU2005} estimate
the upper limit for the electron-phonon coupling constant $\lambda
\approx 0.2$ for Li$_{0.12}$ZrNCl.  A theoretical study on
Li$_x$ZrNCl\cite{HELD2005} predicted the coupling constant on the average
around 0.5. The computed and the estimated coupling constant is far
too small to account for the $T_c$ of 12-15 K. The magnetic
susceptibility measurements \cite{TOU2001} also give a mass enhancement
factor that appears too small for electron-phonon coupling.

There is no clear evidence of strong electronic correlations in the
transition metal nitrides.  These compounds do not show the
antiferromagnetism that is characteristic of strong correlations nor
even the Curie-Weiss susceptibility that signals local moments, and
there is no frustration on either the honeycomb lattice or rectangular
lattice.  In fact, the bandwidths of the $d$ states (where doped
electrons reside) are rather large\cite{WEHT1999} and undoped systems are
in $d^0$ configurations, so these systems should be well described
(except for the value of the gap) by first-principle calculations
employing the local density approximation (LDA).  There is no
observation of magnetism in the parent compounds at all, so the
possibility of spin fluctuations as a pairing mechanism, similar to
what is thought by some to cause superconductivity in cuprates, seems
unlikely.  While a possible pairing mechanism mediated by magnetic
fluctuations has been suggested,\cite{KASA2009} this mechanism seems
at odds with observed behavior so far. Some groups have also proposed
charge fluctuations\cite{BILL2002,BILL2003} as a pairing mechanism or plasmon enhancement of weak BCS superconductivity.

In this paper we focus on the ternary nitride BaHfN$_2$, whose
electronic structure and vibrational properties have not yet been
studied theoretically.  This compound has many chemical and structural
similarities with the layered transition metal nitrides
MNCl's(M=Ti,Hf,Zr) that are impressive superconductors when they are
electron doped.  We suggest that this compound has the potential to
provide another high $T_c$ transition metal nitride superconductor
when electron doped.  In this paper we present calculations of the
electronic structure, lattice vibrations and dielectric constant
tensors of BaHfN$_2$, and compare them with those of other
nitrides. This comparison with other layered nitrides may help in predicting the
origin of the superconductivity in these layered nitrides. 

The rest of the paper is organized as follows. We first describe the
crystal structure of BaHfN$_2$ (Sec.~\ref{sec:structure}) and the
computational methods (Sec~\ref{sec:method}). Then we present our
results for structural and electronic properties of BaHfN$_2$
(Sec.~\ref{sec:electronic}), followed by analysis of the vibrational
properties (Sec.~\ref{sec:vibrational}). Finally, we discuss the case
when BaHfN$_2$ is electron doped by replacing one of the two Ba atoms
in the unit cell by a La atom (Sec.~\ref{sec:doped}). A summary of our
findings in Sec.~\ref{sec:conclusion} concludes the paper.

\section{Structure}\label{sec:structure}

\begin{figure*}[ht]
\begin{center}
\includegraphics[width=14cm]{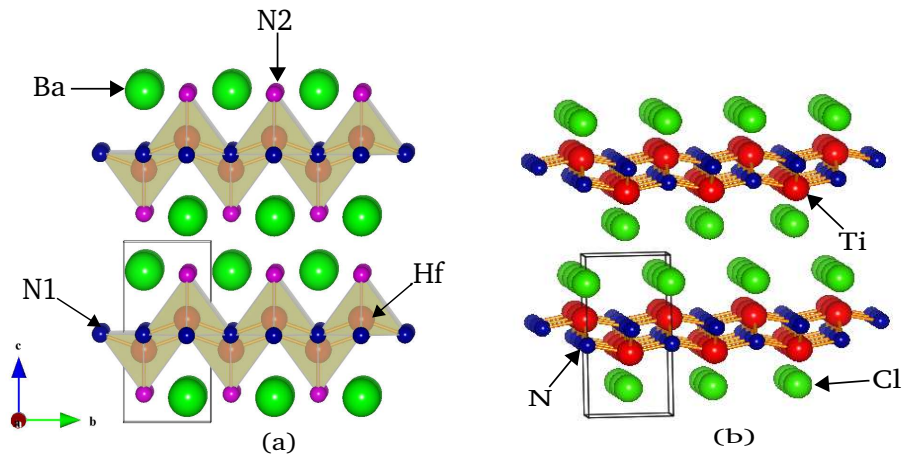}
\caption{(Color online) (a) Layered geometry of BaHfN$_2$: N$^{3-}$
  anions form a square base pyramid around Hf$^{4+}$, forming a
  [HfN$_2$]$^{2-}$ anion.  Ba$^{2+}$ cations sit in between the tips
  of the pyramids.  (b) Layered $\alpha$-TiNCl structure.}
\label{fig:structure} 
\end{center}
\end{figure*}

We use structural coordinates for BaHfN$_2$ from powder X-ray
diffraction measurements reported by Gregory {\it et
  al.}\cite{GREG1998} The nitrodohafnate BaHfN$_2$ crystallizes in
the tetragonal space group P4/nmm, for which KCoO$_2$ is the customary
example.  The measured lattice constants are $a=4.128$~\AA~and
$c=8.382$~\AA.  There are two inequivalent N sites which we denote N1
(lying nearly in the Hf plane) and N2 (nearly in the Ba plane), {\it
  i.e.} BaN2-HfN1.  Ba, Hf and N2 occupy Wyckoff position 2c $(\frac
14,\frac 14, z)$ and N1 occupies position 2b $(\frac 34,\frac 14,\frac
12)$ (see Table I).

The structure of BaHfN$_2$ shown in Fig. \ref{fig:structure} is
sometimes described as composed of [HfN$_2$]$^{2-}$ anions and
Ba$^{2+}$ cations~\cite{GREG1998}. These anions are composed of Hf
atoms inside a square pyramid of 5 nitrogen atoms (four N1's and one
N2), forming layers of edge sharing pyramids stacked along the $c$
axis; the apices of adjoining pyramids are aligned alternatively up
and down along the c axis.  The Ba$^{2+}$ cations are situated between
these Hf-N sheets levelled with the pyramid apices. 

Another description for the structure of BaHfN$_2$ comprises of
(nearly) coplanar (BaN2)$^-$ layer and a corrugated (HfN1)$^+$ layer,
with Hf ions lying alternately above and below the N2 layer.  Each neutral
BaN2-HfN1-N1Hf-N2Ba structural unit (outlined in
Fig. \ref{fig:structure}a) is weakly bonded to neighboring units in
the $c$ direction; upon intercalation, dopant ions will lie between
the Ba-N2 layers. Similarly, in $\alpha$-TiNCl (Fig. \ref{fig:structure}b) there is a
layer of transition metal (Ti) atoms and N atoms, with Cl$^-$ ions
playing a role analogous to the (BaN2)$^-$ unit in
BaHfN$_2$. Neutral TiNCl slabs are
weakly bound to each other. A more detailed study of that compound
will be presented separately.

As we
show below, the site energies and bonding of the N1 and N2 sites are quite different,
and the simple viewpoint of formal closed shell ions, e.g. Ba$^{2+}$,
Hf$^{4+}$, N$^{3-}$ may not be sufficient to address the structural
relationships.

\begin{table*}
\begin{centering}
\caption{\label{tbl:vc-relax}Structurally optimized lattice constants
  ($a$ and $c$) and reduced internal coordinates ($z$) for atoms in
  the unit cell, obtained from different sets of pseudoptentials (PSPs) including Hartwigsen-Goedecker-Hutter (HGH)\cite{HART1998,KRAC2005} and Troullier-Martin (TM)~\cite{TROU1991} type PSPs. 
  The inclusion of semicore $5s$ and $5p$ states for Ba and Hf
  PSPs are indicated by Ba$^{sc}$ and Hf$^{sc}$ respectively. The computed band gap for the experimental geometry, E$_g^{exp}$, is compared with FPLO (FLAPW) results. }
\begin{tabular}{clccccccccc}
\hline\hline
&\multicolumn{1}{c}{PSP} &$a/a_{exp}$&$c/c_{exp}$& $z$(Ba) &$z$(Hf)&$z$(N2)&$E_{g}^{exp}$ (eV) \\ \hline   
LDA &FPLO& 0.992 & 0.986 & 0.846 & 0.415 & 0.177 & 0.68\\
& FLAPW & & & & & & 0.8\\
&HGH, Ba$^{sc}$, Hf$^{sc}$ & 0.986 & 0.982 & 0.848 & 0.415 & 0.176 & 0.82 \\
&TM& 1.005 & 1.017 & 0.846 & 0.415 &0.177 & 0.78\\
&TM,  Ba$^{sc}$ & 0.997 & 0.987 & 0.849 & 0.413 & 0.170 & 0.87 \\
&TM, Ba$^{sc}$, Hf$^{sc}$& 0.984 & 0.972 & 0.847 & 0.415 & 0.177 & 1.11\\
PBE& HGH, Ba$^{sc}$, Hf$^{sc}$ & 1.000 & 1.007 & 0.848 & 0.415 & 0.180 & 0.95 \\
&TM &1.022 & 1.036 & 0.845 & 0.416 &0.186 & 1.00\\
&TM, Ba$^{sc}$  &1.011 & 1.008 & 0.850 & 0.413 & 0.175 & 1.13 \\
&TM, Ba$^{sc}$, Hf$^{sc}$&1.004 & 0.999 & 0.849 & 0.414 & 0.179 & 1.25\\
 Exp.\cite{GREG1998}  & & & & 0.8479 & 0.4142 & 0.168 & \\
\hline
\end{tabular}
\end{centering}
\end{table*}

\section{Description of Calculations}\label{sec:method}
We carried out density functional theory (DFT) calculations with the
ABINIT package~\cite{abinit}, within both local density approximation
(LDA) and gradient corrected (GGA/PBE~\cite{PERD1996}) exchange and
correlation functionals. Norm-conserving pseudopotentials (PSPs) in the
relativistic separable dual-space Gaussian
Hartwigsen-Goedecker-Hutter (HGH) form\cite{HART1998,KRAC2005} were
used to treat the electronic configuration of Ba (5s, 5p, 6s), Hf (5s,
5p, 5d, 6s) and N (2s, 2p), including $5s$ and $5p$ semicore states
for Ba and Hf. Plane wave basis sets with a kinetic energy cutoff of
120 Ry were used. A $8\times 8\times 4$ Monkorst-Pack \cite{MONK1976}
$k$-point grids were used to sample the Brillouin zone for ground
state calculations. We have checked that further increasing the cutoff
energy to 140 Ry or the $k$-point grid to $12 \times 12 \times 6$ and
$18 \times 18 \times 6 $ has a negligible influence on the relaxed
geometry and phonon frequencies. The computed LDA band structure of
BaHfN$_2$ at experimental geometry was found to agree well with
results obtained from the full-potential, all electron code
FPLO~\cite{FPLO}; the latter also provides a convenient way to compute
contributions to the electronic bands and density of states from
individual atomic orbitals. We have confirmed that spin-orbit coupling
has no significant influence to the band structure. Phonon
calculations are carried out at the $\Gamma$ point, and the obtained
frequencies and displacement eigenmodes were used to compute Born
effective charges and the static dielectric tensor $\epsilon_0$.

To examine the influence of PSPs on the calculated
structural, electronic and vibrational properties of BaHfN$_2$,
especially the inclusion of $5s$, $5p$ semicore states for Ba and Hf,
we also carried DFT calculations using Troullier-Martin
(TM)~\cite{TROU1991} PSPs generated using the fhi98PP
program~\cite{FUCH1999} with LDA/PZ ~\cite{PERD1981} and GGA/PBE
exchange-correlation functionals, respectively.  A $8\times 8\times
4$ $k$-grid and a kinetic energy cut off of 90 Ry were used. 

To simulate the doped BaHfN$_2$, we replaced one of the two Ba atoms
in the unit cell with La, which provides an extra electron per formula 
unit and provides metallic screening with its impact on the 
zone-center phonons. The HGH PSPs were used to do calculations for the doped system with kinetic energy cut-off of 120 Ry and and a k-point mesh of $12\times 12\times 6$. 


\section{Electronic structure Calculations}\label{sec:electronic}
\subsection{Structural Relaxation and Electronic Structure}

It is instructive for future studies to quantify the effects of
different types of pseudopotentials on the relaxed structure and the
energy gap. 
We have performed structural relaxation for the BaHfN$_2$ crystal using
both LDA and GGA/PBE, and results are compared with experimental
geometry~\cite{GREG1998} in Table~\ref{tbl:vc-relax}.  In addition to
the HGH PSPs, we also employed different sets of
norm-conserving TM PSPs. The inclusion of $5s$ and $5p$ semicore
states, if present, is denoted by Ba$^{sc}$ and Hf$^{sc}$
respectively. The optimized cell parameters and internal coordinates
were compared with all-electron, full potential reference results from
FPLO.\cite{FPLO} The band gap was computed for the
experimental geometry, $E_g^\mathrm{exp}$ using FPLO~\cite{FPLO} and 
full potential linear augmented plane waves code (FP-LAPW) implemented 
in the Elk code.\cite{elk}  The results are listed in
Table~\ref{tbl:vc-relax}.

\begin{figure}[tbp]
\includegraphics[width=\columnwidth]{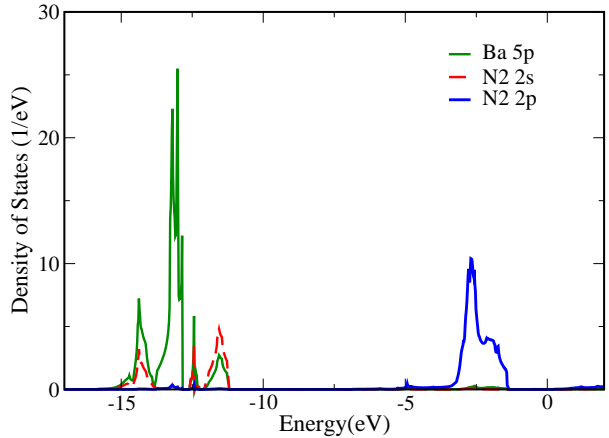}
\caption{(Color online) Projected density of States(PDOS) showing strong hybridization of Ba 5p semi-core states with N2 2s states and a weak mixing with N2 2p states.}
\label{fig:Basemicore} 
\end{figure}

One observation from Table~\ref{tbl:vc-relax} is that as one includes
semicore states of Ba and Hf, the equilibrium lattice constants
becomes smaller, while the energy gap $E_{g}^{exp}$ increases
substantially. Such trend holds for both LDA and PBE PSPs of TM type.
The inclusion of semi-core states is important since there is
significant amount of hybridization of Ba semi-core $5p$ states with N2
$2s$ state, as can be seen from Fig.~\ref{fig:Basemicore}. We observed a trend of decrease in equilibrium lattice constants upon inclusion of semi-core states. This is mainly due to the decrease of Ba-N2 and Ba-Hf bond length.

On the other hand, TM-type PSPs generated using the fhi98PP code are
so-called single projector pseudopotentials, e.g. there is only one
pseudopotential for each angular momentum type, not for each valence
orbital. The corresponding PSPs with Ba$^{sc}$ and/or Hf$^{sc}$ failed
to describe the energy position of 6s states properly, even for the
isolated atoms. Such PSPs tend to predict energy gaps of BaHfN$_2$
larger than those without the semicore stats in the valence
configuration. The discrepancy in $E_g^\mathrm{exp}$ compared with
all-electron calculations can be as much as 0.4 eV. The HGH
pseudopotentials, on the other hand, were constructed with multiple
projectors per angular momentum type and therefore can describe
orbitals of same angular momentum but different shells reasonably well.
Indeed, from Table~\ref{tbl:vc-relax}, we find that overall HGH
pseudopotentials give structural properties and energy gap similar to
those calculated from all-electron calculation and in the following,
we present results obtained with LDA type HGH
pseudopotentials.

\begin{figure}[t]
\begin{center}
\includegraphics[width=\columnwidth]{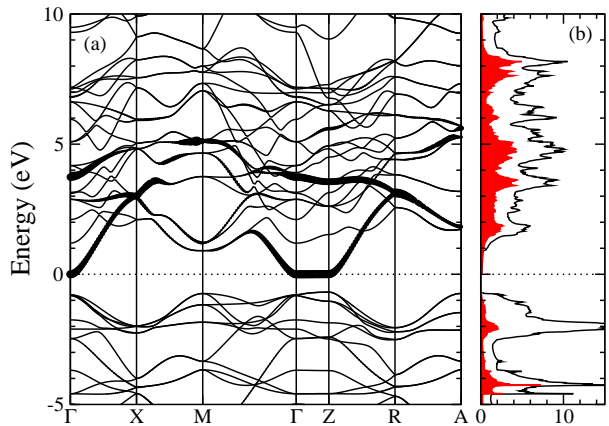}
\caption{(a) Band structure showing Hf 5d bands.  The amount of Hf
  $5d_{xy}$ character is shown in the band structure by the width of
  the lines, revealing that the lowest conduction band
  has strong Hf $5d_{xy}$ character.  (b) The density of states, units for
  horizontal axis are eV$^{-1}$.  The shaded area indicates the amount
  of Hf $5d$ character. The Fermi level lies at the bottom of the lowest conduction band.}
\label{fig:Hf 5d bands}
\end{center}
\end{figure}

Figure \ref{fig:Hf 5d bands} shows the band structure of BaHfN$_2$, using the so-called fatbands emphasis of band character for Hf $5d_{xy}$ states.  The
fatbands are obtained by using the expansion of the wavefunctions in
terms of the basis atomic orbitals at each k-point
\begin{equation}
|\mathbf{k}n\rangle =  
		\sum_{\mathbf{Rs}L} c_{Ls}^{\mathbf{k}n} 
		e^{i\mathbf{k}\cdot(\mathbf{R}+\mathbf{s})} |\mathbf{Rs}L\rangle, 
\end{equation}
\noindent where $n$ is the band index and $L \equiv lm$ is the orbital
index. $\mathbf{R+s}$ denotes a regular lattice site, with
$\mathbf{R}$ a Bravais lattice vector and $\mathbf{s}$ a basis vector
of the unit cell.  The width of the fatband is proportional to
$|c_{L\mathbf{s}}^{\mathbf{k}n}|^2$. 

BaHfN$_2$ is a band insulator with a calculated band gap
0.68-0.80 eV using full-potential, all-electron methods (FPLO, LAPW). 
Given the usual LDA underestimate of band gaps the true gap of BaHfN$_2$ may be as large as 1.5 eV.  This
layered ionic semiconductor character is very similar to that of the
MNCl compounds (M=Ti, Hf, Zr) which have been found to superconduct
with impressively high $T_c$ values when electron-doped.  The lowest
conduction band in BaHfN$_2$ has primarily Hf $5d_{xy}$ character
with a width of 3 eV. Since these states are empty, Hf is formally 4+, and
the rest of the electronic structure is indicative of a closed shell,
ionic insulator with some mixing of N $2p$ states and Hf $5d$ states.
This characterization is similar to ZrNCl, which is also an ionic
semiconductor with lowest conduction band having Zr in-plane $4d$
character,\cite{WEHT1999} and TiNCl which has Ti $3d_{xy}$ character
[\onlinecite{YAMA2009},\onlinecite{Quan-TiNCl}].  The Hf $5d$
character extends through a range of 8 eV beyond the Fermi level (see
Fig. 2b) partially due to crystal field splitting of the $5d$
orbitals.

\begin{figure}[t]
\includegraphics[clip,width=\columnwidth]{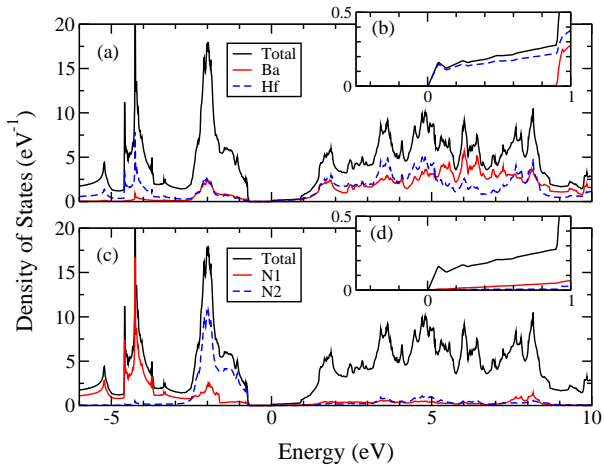}
\caption{(Color online) Total density of states (DOS) for BaHfN$_2$,
  with projected density of states for (a) Ba, Hf with (b) blowup of
  conduction bands, and (c) N1, N2 with (d) blowup of conduction
  bands.  The lowest conduction bands, are composed of Hf $5d_{xy}$
  states.  (b) and (d) show almost 2D like behavior around the bottom
  of the conduction band up to $\sim$0.9 eV, where Ba conduction
  states start to appear.  }
\label{fig:DOS}
\end{figure}

Figure \ref{fig:DOS} shows the projected density of states in the
valence-conduction band region.
Integrating the density of states, we find that only above a doping
level of 0.17 electrons (shifting the Fermi level upward by 0.9 eV) do
bands other than the two-dimensional Hf $5d_{xy}$ band start filling up with electrons.  The conduction bands which appear at 0.9 eV above the Fermi
level are Ba $5d$ states at M that have some dispersion along k$_z$
(compare the points M and A).
There is also some nitrogen hybridization in the conduction bands that
contribute to their in-plane dispersion.

As mentioned earlier, there are two N sites, corresponding to the
corrugated layers Hf$^{4+}$N1$^{3-}$ and the nearly flat Ba$^{2+}$N2$^{3-}$ layers.  The projected
DOS shows that the N1 and N2 ions are quite distinct in electronic
character.  The N2 ion in the BaN2 layer has the more weakly bound
(and therefore more polarizable) N $2p$ states, lying just below the
gap.  The energy states of N1 ion in the HfN1 layer (closer to the highly charged Hf$^{4+}$
ion) is centered about 2 eV lower in
energy. From purely energetic (binding) consideration, N1 should have
a correspondingly lower polarizability.


\subsection{Fermi Surface for Electron Doping}
According to our band structure calculations in Fig.\ref{fig:Hf 5d bands},
when BaHfN$_2$ is electron-doped (as are ZrNCl, HfNCl, and TiNCl when
they become superconducting), the Fermi surface will be a single $\Gamma$-centered
(nearly) circular surface up to $x$=0.17 for doping concentration $x$ when doped with alkali metals; for doping concentration x $>$ 0.17 carriers also go into the bottom of the Ba $5d$ bands at M and
then additional Hf $5d$ bands at $\Gamma$.  In this respect, BaHfN$_2$
is similar to TiNCl\cite{Quan-TiNCl}, but different from the hexagonal
compounds [(Zr,Hf)NCl] which have the band minimum at the zone
corner K points, of which there are two.

This difference in Fermi surfaces has some importance for electronic response.  In the doped
(Zr,Hf)NCl compounds, nesting of the two Fermi surfaces has recently
been put forward\cite{KURO2010} as a potential source of spin
fluctuations, which was suggested as a possible candidate for pairing
mechanism.  However, with a single simple Fermi surface such as
displayed by doped TiNCl (known to be an excellent superconductor) and
doped BaHfN$_2$ (which we suggest by analogy may be a good
superconductor), this mechanism is not available.  Since the
superconducting T$_c$ is large (and almost similar in magnitude) in TiNCl and ZrNCl,
and their characters are otherwise so similar but Fermi surfaces are
different, the mechanism of spin fluctuations seems to be degraded in
likelihood. Electron-phonon coupling is weak in A$_x$ZrNCl, and the materials 
are pauli-paramagnetic.  Thus for possible pairing in these materials,
long sought electronic mechanisms of pairing need consideration. A single electron gas with a given value of $k_F$ is different from a pair of identical, degenerate electron gases with a value of $k_F$ that is 1/$\sqrt{2}$ as
large. One main difference, as mentioned above, is that there is no nesting that 
might enhance charge fluctuations (as is the case also for
spin fluctuations) or affect pairing symmetry. Another clear difference is that the
characteristic momentum scale $k_F$ is different.

\section{Vibrational Spectrum}\label{sec:vibrational}

\subsection{Born-Effective Charges}

\begin{table*}

\begin{centering}
\caption[Effective Charges]{Born effective charges (BECs) of BaHfN$_2$
  and metallochloronitrides MNCl (M=Hf,Zr,Ti). All BECs have been
  calculated at experimental lattice constants and relaxed atomic positions.  $\alpha$-TiNCl is orthorhombic, so $Z^*_{xx} \neq Z^*_{yy}$ and both are listed.  For comparison, BECs of NaCoO$_2$~\cite{LI2004} are also listed. }
\label{tbl:BEC}
\begin{tabular}{r|cccc|ccc|ccc|ccc|ccc}
\hline
\hline & \multicolumn{4}{c|}{BaHfN$_2$} &
\multicolumn{3}{c|}{$\beta$-HfNCl} &\multicolumn{3}{c|}{$\beta$-ZrNCl}
& \multicolumn{3}{c|}{$\alpha$-TiNCl}
&\multicolumn{3}{c}{NaCoO$_2$}\\ \hline
& Hf & N1 & N2 &Ba & Hf & N & Cl & Zr
& N & Cl & Ti & N & Cl & Na & Co & O\\ 

$Z^*_{xx}$/$Z^*_{yy}$ & 4.52 & -4.66 & -2.59 &2.73 & 4.7 & -3.4 & -1.2
& 5.1 & -3.8 &-1.3 &5.9/6.4 & -5.5/-4.4 &-0.4/-1.9 &0.87 &2.49
&-1.68\\

 $Z^*_{zz}$ & 3.09 & -1.65 & -4.58 &3.14 & 2.5 & -1.6 & -0.9 &2.6 &
-1.6 & -1.0 & 1.7 & -0.9 & -0.8 &1.37 &0.87 &-1.12 \\
 $Z^{\text{Formal}}$  &   +4   & -3    & -3     & +2  & +4  & -3   &-1  &+4 & -3 & -1  &+4 & -3 & -1 &+1 &+3 &-2\\
\hline
\end{tabular}

\end{centering}
\end{table*} 

The MNCl compounds become superconducting upon electron doping from an
ionic insulator to the metallic phase.  The relatively
low density electron gas that is formed upon light doping might not adequately screen the ionic nature of the MN layers, so the electronic 
response may still have short-range ionic character.  For this reason we
calculate and analyze the Born effective charges in some detail.

The Born effective charge (BEC) tensor $\mathbf{Z}^{*}$ is a
fundamental quantity for the study of lattice dynamics, describing the
long range Coulomb part of the force constants. The Born effective
charge $Z_{\kappa,\gamma \alpha}^{*}$ of atom $\kappa$ can be viewed
either as the change of polarization $P_{\gamma}$ induced by the
periodic displacement $\tau_{\kappa,\alpha}$ under the condition of
zero macroscopic electric field, or as the force $F_{\kappa,\alpha}$
induced on atom $\kappa$ by an electric field $\cal{E}_{\gamma}$ under
the condition of no atomic displacement.  It can also be expressed as
the second partial derivative of the total energy with respect to the
displacement and the electric field:
\begin{equation}
Z_{\kappa,\gamma \alpha}^{*}
=  V \frac{\partial P_{\gamma}}{\partial \tau_{\kappa,\alpha}}
=  \frac{\partial F_{\kappa,\alpha}}{\partial \cal{E}_{\gamma}}
= - \frac{\partial^{2} E}{\partial \cal{E}_{\gamma}
 \partial \tau_{\kappa,\alpha}}
\end{equation}
where V is the volume of the unit cell.  One might naively expect the
BECs to be close to the formal charges of the compound, but this is
often not the case. The BEC can be decomposed into the charge of the
(pseudo-) ion $\kappa$, Z$_{\kappa}$, and the electronic screening
term, $\Delta Z_{\kappa,\gamma \alpha}$.
\begin{equation}
Z_{\kappa,\gamma \alpha}^{*}=Z_{\kappa,\gamma \alpha} + \Delta
Z_{\kappa,\gamma \alpha}
\end{equation}

The computed BECs for BaHfN$_2$ are provided in Table \ref{tbl:BEC}.
In tetragonal symmetry the BEC tensor is diagonal and reduces to two
values $Z_{xx}^{*}=Z_{yy}^{*}$ and $Z_{zz}^{*}$.  The BECs for Hf and
Ba in the plane are reasonably close to their formal
charges (Hf is 0.52 larger and Ba is 0.73 larger), while their perpendicular charges differ substantially from
the formal charges (being smaller than the formal charge for Hf)
indicating a more complex electronic response.

The two nitrogen sites have very different and unusual BECs.  For
comparison, previously calculated\cite{LI2004} BECs of NaCoO$_2$ are
also listed in Table \ref{tbl:BEC}.  The BECs for O in NaCoO$_2$ are
rather uninteresting, both being smaller than their
respective formal charges.  By contrast, in BaHfN$_2$, N1 has a BEC of
-4.66 in the plane and -1.65 perpendicular to the plane,
respectively with magnitude much larger and much smaller than the
formal charge.  N2, on the other hand, has a BEC of -2.6 in the plane
and about -4.6 perpendicular to the plane, again very different from
the formal charge but {\it in the opposite sense} with respect to N1.
The BEC of N1 is consistent with covalent bonding between N1 and Hf,
given its anomalously large magnitude in the plane.  N2 behaves in the
opposite way, and its BEC is consistent with little covalent
bonding with Ba (as expected) but with significant covalent
interaction with Hf in the inner layer. BECs with magnitudes greater than the formal charges reflect large electronic response to atomic motion. For
example, in the case of perovskite BaHfO$_3$\cite{VALI2008} the large
Born-effective charges of Hf (Z$^*$ = 5.75 ) and O ( Z$^*_{O||}$ =
-4.42, Z$^*_{O\perp}$ = -2.03) indicate a mixed ionic-covalent nature
of Hf-O bond, similar to the case found here for the Hf-N1 bond. Also, Ba
in BaHfO$_3$, which has a cubic site symmetry, was found to have a
similar average BEC (Z$^*$ = 2.72 as that computed for Ba in
BaHfN$_2$ here ( Z$^*$ = 2.87)).

The Hf-N1 layer, taken as a unit, behaves as if having a charge of
-0.14 (nearly neutral) in the plane and +1.44 (slightly cationic,
rather consistent with the formal charges) for vibrations
perpendicular to the plane. The Ba-N2 layer behaves in an opposite
manner.

In Table~\ref{tbl:BEC}, we draw a comparison between the BECs of
metallochloronitrides MNCl (M=Ti,Hf,Zr)\cite{BEC} and those of
MNCl (M=Ti, Hf, Zr)\cite{Quan-TiNCl} show similar trends
as do the BECs of N1 in BaHfN$_2$ and the N in MNCl.  There
is considerable anisotropy in the effective charges for both the M and
N ions.  The effective charge for Cl in MNCl is close to its
formal charge (reflecting its high electronegitivity), however the Cl
analog in BaHfN$_2$, Ba-N2, is somewhat different, with large
anisotropy. 

\subsection{Zone-Center Phonons}

\begin{table*}[bht]
\begin{centering}
\caption[Zone-center phonon frequencies] {Calculated zone-center
  phonon frequencies for BaHfN$_2$ and Ba$_{0.5}$La$_{0.5}$HfN$_{2}$. 
  The phonons were computed 
  using the optimized geometry. All phonons with $x$-$y$ polarization are
  doubly-degenerate.  The symmetry column refers to the symmetry of the 
  phonons in higher symmetry insulating system 
  (point group $D_{4h}$ vs. $C_{4v}$ for the
  metallic system).  There is a split in 
  degeneracy in the long wavelength limit between
  the LO and TO modes, and the magnitude of LO-TO splitting,
  $\sqrt{\omega^{2}_{LO}-\omega^{2}_{TO}}$ , is listed in the last
  column.  All frequencies reported in cm$^{-1}$.}
\begin{tabular}{rcccccc}
\hline\hline Mode & Symmetry & Polarization &\multicolumn{3}{c}{BaHfN$_2$ frequency}   & Ba$_{0.5}$La$_{0.5}$HfN$_2$ \\
\footnotetext[1]{IR active mode}
\footnotetext[2]{Raman active mode}
&  & & $\omega_{TO}$ & $\omega_{LO}$  & $\sqrt{\omega^{2}_{LO}-\omega^{2}_{TO}}$ & frequency \\
\hline
1-2\footnotemark[1]           & E$_{u}$     &   $x$-$y$ &   $\mathbf{72}$    &$\mathbf{93}$  & $\mathbf{59}$     & 76 \\
3-4\footnotemark[2]     & E$_{g}$   &   $x$-$y$ &   82    &  &  & 94 \\
5\footnotemark[1]   & A$_{2u}$  &  $z$  &   $\mathbf{105}$ & $\mathbf{144}$ & $\mathbf{98}$ & 144\\
6\footnotemark[2]   & A$_{1g}$  &$z$&   120 &    & & 136\\
7-8\footnotemark[2] & E$_{g}$   &   $x$-$y$ &   152 &  & & 148 \\
 9\footnotemark[2]   & A$_{1g}$     &  $z$  &   172 &   & & 175 \\
10-11\footnotemark[1]  & E$_{u}$    &   $x$-$y$ &   $\mathbf{210}$ & $\mathbf{240}$ & $\mathbf{116}$ & 210\\
12-13\footnotemark[2]  & E$_{g}$   &    $x$-$y$ &   232 &   & & 283\\
14\footnotemark[2]  & B$_{1g}$  &  $z$  &   341 &   &   & 328 \\
15-16\footnotemark[1]  & E$_{u}$  &   $x$-$y$   &   $\mathbf{424}$ & $\mathbf{614}$ & $\mathbf{444}$ & 475\\
17\footnotemark[1]  & A$_{2u}$   &   $z$&   $\mathbf{468}$  & $\mathbf{492}$ &  $\mathbf{152}$  & 457\\
18-19\footnotemark[2]  & E$_{g}$  & $x$-$y$ &  623 &  &   & 651 \\
20\footnotemark[1]  & A$_{2u}$ &   $z$ & $\mathbf{641}$  & $\mathbf{751}$  & $\mathbf{391}$   & 596 \\
21\footnotemark[2]  & A$_{1g}$  &  $z$  &   717  &   &   & 646 \\
\hline
\end{tabular}
\label{tbl:PhononFrequencies}
\end{centering}
\end{table*}

\begin{table}[b]
\begin{centering}
\caption{Calculated macroscopic dielectric constants for BaHfN$_2$ and 
  group IVB nitrochlorides using (a) fully relaxed geometry and (b)
  the experimental structure with relaxed atomic positions.}
\label{tbl:Dielectric}
\begin{tabular}{rrcccc}
\hline\hline && $\epsilon^{\infty}_{xx/yy}$&$\epsilon^{\infty}_{zz}$ &
$\epsilon^{0}_{xx/yy}$&$\epsilon^{0}_{zz}$\\\hline
(a)& BaHfN$_2$  & 7.47 & 7.55 & 33.8 & 21.4 \\
(b)& BaHfN$_2$ & 7.35 &7.31 & 44.7 & 24.0\\ 
(b)&$\alpha$-TiNCl &6.9/7.4 & 3.2 & 22.3/38.3 &3.7\\
(b)&$\beta$-ZrNCl & 6.2 & 4.4 & 13.8 &5.9\\  
(b)&$\beta$-HfNCl & 5.4 & 4.0 & 11.1 & 5.1\\ \hline 
\end{tabular}
\end{centering}
\end{table}

\begin{table}[b]
\begin{centering}
\caption{Lattice contribution (defined in Eq. \ref{eq:ModeContrib}) to the macroscopic dielectric constants for selected phonons computed using fully relaxed geometry.  Only 3 phonon modes contribute to the lattice polarizability in each direction. }
\label{tbl:Oscillator}
\begin{tabular}{rc|c|c|c}
\hline\hline & Mode & $\omega_{TO}$ (cm$^{-1}$) & 
$4\pi P_{m,xx}$ & $4\pi P_{m,zz}$\\\hline
&1-2 & 72 & 14.0 & \\ 
& 5  & 105 & & 10.1\\
& 10-11 & 210 & 5.1 & \\  
& 15-16 & 424 & 7.3 &\\  
& 17 &468 & & 1.3 \\ 
& 20 &641 & & 2.4\\ \hline
\end{tabular}
\end{centering}
\end{table}

BaHfN$_2$ has 8 atoms in the unit cell resulting in 24 phonon modes,
three of which are acoustic modes and the remaining 21 are optical
modes.  The phonon frequencies are listed in Table
\ref{tbl:PhononFrequencies} with their polarization and symmetry.
There are $8E_g + 6E_u $ modes with polarization perpendicular to the
$c$ axis (within the $x$-$y$ plane) and $3A_{1g} + 3A_{2u} + 1B_{1g}$
modes with polarization along the $z$ axis. The modes A$_{1g}$,
B$_{1g}$, E$_g$ are Raman active, and the modes $A_{2u}$, E$_u$ are
infrared active.
We found some phonon frequencies were quite sensitive to 
the inclusion of semi-core states in Ba and Hf pseudopotentials with
several phonons differing by 15-30\% in the absence of semi-core states.

The LO-TO splittings of the IR active modes can be
related to the Born-effective charges as
\begin{equation}
\sum_m{[\omega^2_{LO,m}-\omega^2_{TO,m}]}=\frac{4\pi}{\epsilon{_{\alpha\alpha}^\infty}V}\sum_\kappa{\frac{{(eZ{_{\kappa,\alpha\alpha}^*})}^2}{M_\kappa}}.
\end{equation}
\noindent In this relation, $m$ goes over the IR active modes of a
given polarization direction $\alpha$, $M_\kappa$ is the ionic mass of
the atom $\kappa$, and $\epsilon_{\alpha\alpha}^\infty$ is the
$\alpha$-th diagonal element of the high frequency dielectric
constant.  When the LO phonons are excited, a macroscopic electric
field is created due to the long range nature of the Coulomb
interaction. 
The squares of the BECs, divided by the mass, give the 
contribution of that ion to the electric field.  One interesting example (see Table \ref{tbl:PhononFrequencies}): the large
splitting ($\Delta\omega$) of mode 15 is due largely to the fact
that the light N1 ions are vibrating in the $x$-$y$ plane and the BEC for N1 is
rather large (-4.66), accounting for most of the shift of 45\%
in LO frequency. The second largest splitting ($\Delta\omega$) is for the
mode 20, with a shift of about 17\% in frequency. In this mode we have
primarily N2 vibrating along the $z$ direction and the BEC for N2
along $z$ is $Z^{*}_ {zz}$(N2) = -4.58 which again accounts for the
large splitting. Mode 5 has a large relative shift of 37\%; it has
primarily N1 and Ba vibrations opposite to each other.


\subsection{Dielectric Response}
We now discuss the electronic and ionic lattice contributions to the macroscopic 
dielectric constants, computed for BaHfN$_2$ by doing the phonon calculation 
using both experimental lattice constants with relaxed geometry and completely relaxed geometry. $\epsilon^{\infty}$ denotes the high 
frequency electronic response where there is no contribution from the ionic lattice polarizability ($P^{ion}$)
and $\epsilon^{0}$ is the sum of the electronic
and ionic response.  Their relationship is given by
\begin{eqnarray}
\epsilon_{\alpha\beta}^{0}=\epsilon^{el} + 4\pi P^{ion} &=&
        \epsilon_{\alpha\beta}^{\infty} + 4\pi \sum_m P_{m,\alpha\beta} \\
P_{m,\alpha\beta} &=& \frac{1}{\Omega}\frac{S_{m,\alpha\beta}}{\omega_m^{2}},
\label{eq:ModeContrib}
\end{eqnarray}
\noindent where the sum is over all the modes and $S_{m,\alpha\beta}$
is the mode-oscillator strength tensor which is defined as
\begin{equation}
S_{m,\alpha\beta}=\sum_{\kappa,\alpha'}{Z_{\kappa,\alpha\alpha'}^{*}U_m(\kappa,\alpha')}
\times
\sum_{\kappa',\beta'}{Z_{\kappa',\beta\beta'}^{*}U_m(\kappa',\beta')},
\end{equation}
\noindent where $U_m(\kappa,\alpha)$ is the component of the phonon
eigenvector for the $m$-th mode corresponding to the displacement of
atom $\kappa$ in direction $\alpha$.

The values of the static dielectric constants for BaHfN$_2$ and comparison to
the group IVB metallochloronitrides\cite{BEC} are given in Table \ref{tbl:Dielectric}. 
$\epsilon^\infty$ for BaHfN$_2$ is larger
than for the metallochloronitrides (the electronic polarizability $\epsilon^\infty$-1
is 50-60\% larger than the Hf counterpart), consistent with the
smaller band gap (0.8 eV versus around 1.8 eV for ZrNCl and HfNCl).

With the exception of the in-plane values for TiNCl, the lattice polarizability
$\epsilon^0 - \epsilon^\infty$ of other MNCl's is smaller 
by a factor of 5-15 relative to that of BaHfN$_2$. This is due to the fact that the phonons in BaHfN$_2$  are softer than the phonons in MNCl's and the oscillator strengths for modes which contribute to $P^{ion}$ in BaHfN$_2$ are much larger than the modes contributing $P^{ion}$ for the other chloronitrides.
The modes that contribute to lattice polarizability $P^{ion}$ in 
BaHfN$_2$ are shown in Table \ref{tbl:Oscillator} together with their contributions.  The main contribution 
arises from the lowest IR active mode for each direction ($x$ and $z$).
We found that the frequency of the lowest IR mode in the plane is highly sensitive to the inclusion of the semi-core states in Ba pseudopotential, and
as a result when these states are absent $\varepsilon^0_{xx}$ is reduced from 
44.7 to 17.3.

For some comparison we note the dielectric constants for a few transition metal nitrides. The high frequency
dielectric constants for group IVB nitrides have been
reported\cite{XU2006} as follows : Ti$_3$N$_4$, Zr$_3$N$_4$,
Hf$_3$N$_4$ with $\epsilon^\infty$ = 18.31, 9.36, 10.10
respectively. Ti$_3$N$_4$ has a higher dielectric constant due to a
small band gap. 

\section{Doping with electrons}\label{sec:doped}

\begin{figure}[t]
\begin{center}
\includegraphics[width=\columnwidth]{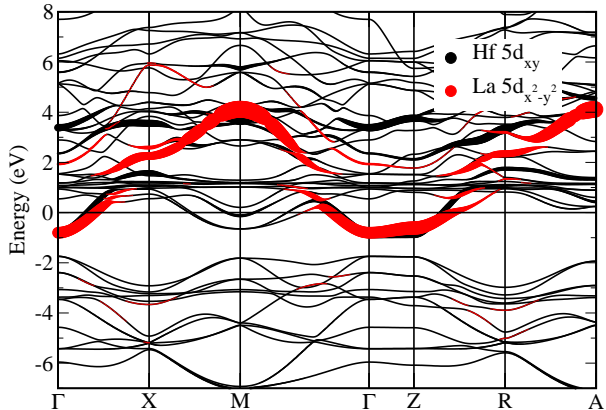}
\caption{Doped band structure showing La and Hf 5d bands.  The amount of Hf
  $5d_{xy}$ and La $5d_{x^2-y^2}$ character is shown in the band structure by the weight of the points in the lines. The lowering of the 5$d$ character due to replacement of Ba by La is substantial. The Fermi level lies at an energy of 0 eV.}
\label{fig:Doped}
\end{center}
\end{figure}

We consider La substitutional doping by replacing 
one Ba in the unit cell with La. 
The vibrational frequencies for the doped system
are included in Table \ref{tbl:PhononFrequencies}.  Since there
is no experimental data on Ba$_{0.5}$La$_{0.5}$HfN$_2$ we use
relaxed lattice constants, which are smaller by almost 3\% than the 
experimental lattice constants of BaHfN$_2$. 
The band structure of (BaLa)$_{0.5}$HfN$_2$ is shown in Fig.~\ref{fig:Doped} 
with fat bands for Hf and La.  
The nearly dispersionless La 4f states are located 1 eV above $E_F$ but have no clear impact on what
we discuss in the following.
There are two nearly cylindrical Fermi surfaces 
around $\Gamma$, one of which has Hf $5d_{xy}$ character and another has La $5d_{x^2-y^2}$ 
character; mixing may occur at or near crossing of the Fermi surfaces.
The lowest conduction bands near M, primarily La $5d_{yz}$, $5d_{xz}$ in character (not
shown as fatbands) are significantly lowered from the corresponding undoped Ba  
bands, creating two Fermi 
surfaces that are larger than anticipated from a rigid band picture using the
BaHfN$_2$ bands.  These surfaces have significant three dimensional character, but even at this (large) doping 
level they do not reach the top of the zone (the A point).  The lowering of the bands 
around M is also seen within the virtual crystal approximation (where both Ba and La are replaced
by an `average ion') so we expect this to be a robust feature for doping by La. 
A small Fermi surface near M, of mainly La $5d_{z^2}$ character, arises near this level of doping. 

The phonon frequencies for the metallic system are shown in Table \ref{tbl:PhononFrequencies}. 
Although the replacement of one Ba with La changes the symmetry, corresponding
modes between the two systems can be identified by examining 
the scalar products of their eigenvectors.
Several of the softer modes
have slightly higher frequencies in (BaLa)$_{0.5}$HfN$_2$
likely due to the decreased lattice constant of the metallic system.
More interestingly, several of the high frequency modes are renormalized
in the metallic system.  These high frequency modes are dominated
by motion of the N1 and N2 atoms.  

\section{Summary}\label{sec:conclusion}

We have examined the electronic and vibrational structure
of the ternary nitride BaHfN$_2$ within density functional theory.  We
find that BaHfN$_2$ seems to have chemical and electronic
similarities with high $T_c$ metallochloronitrides MNCl's (M=Ti, Hf,
Zr), so its candidacy as another high $T_c$ superconducting nitride is
plausible.  The basic electronic and vibrational properties of the
undoped insulating phase provide a basis for an understanding of the
behavior of BaHfN$_2$ upon doping.  We find highly anisotropic Born
effective charges for the N ions, with anisotropies that have an
opposite sign for the two N sites.  These differences suggest unusual electronic screening in BaHfN$_2$ , so that in the absence of strong correlations
and magnetism, electron-electron interactions might play a significant
role in pairing in these layered nitrides when electron doped.  The
large BECs result in large LO-TO splittings for some zone center
phonons, as well as large dielectric constants that also imply unusual
characteristics of electronic screening. We also provided initial analysis
of how the system is affected by doping and found that
bands near M are significantly lowered so that conduction would occur
both in the Hf states near $\Gamma$ and in states near M located in the BaN layer.

The MNCl compounds, which are impressive superconductors when doped, provide an interesting analogy to BaHfN$_2$, with
its similar structural, electronic, and vibrational similarities but larger electronic screening.  One
potentially important difference is worth noting.  From the point of
view of vapor phase growth of the materials, MNCl contain two
reactive, highly electronegative anions.  There is relatively little
experience in vapor growth (molecular beam epitaxy, pulsed laser
deposition) of such materials.  Since the appearance of the high T$_c$
cuprates, there has been a huge amount of experience accumulated, and expertise
gained, in deposition of oxides with one electronegative anion but
several cations. From this viewpoint, BaHfN$_2$ -- with one anion and
two cations -- is attractive for vapor phase growth, and hence for
study and potential application of ultra-thin superconducting layers.

\section{Acknowledgments}

We acknowledge helpful conversations with Dr. Quan Yin, particularly
on unpublished work on TiNCl. We also acknowledge J. N. Eckstein for his useful comments in the early stages of this work. This work was supported by DOE/SciDAC
Grant No. DE-FC02-06ER25794.

\appendix
\section{Description of the vibrational modes}

Here we provide a brief characterization of the eigenvectors of all
optical modes at $q=0$, which can be useful in interpreting optical
data and in comparing with similar ionic semiconductors.  The units
are cm$^{-1}$.

 $\omega = 72$ : Ba oscillating against other atoms in the $x$-$y$
plane. These two degenerate soft mode is largely responsible for the
large static dielectric constant $\epsilon^0_{xx/yy}$.

 $\omega = 82$ : This mode is primarily out-of-phase Ba vibrations in
the $x$-$y$ plane.

 $\omega = 105$ : Ba and N1 moving against each other with strong
amplitude along the $z$ axis.  Hf and N2 are in phase with each other
oscillating weakly as compared to Ba and N1.

 $\omega = 120$ : This mode is primarily out-of-phase Ba vibrations
along the $z$ axis.

 $\omega= 152$ : This mode has primarily N2 and Hf vibrations in the $x$-$y$ plane with N2 vibrating with a large amplitude as compared to Hf with Ba and N1 participating very weakly.

 $\omega = 172$ : Hf and N2 vibrating along $z$, with Ba vibrating
very weakly opposite to Hf.

 $\omega = 210$ : These two mode are primarily in-phase N2 vibrations
in the $x$-$y$ plane.

 $\omega = 232$ : These two modes are primarily out-of-phase N2
vibrations in the $x$-$y$ plane.

 $\omega = 341$ : This mode exhibits pure N1 vibrations along the $z$
axis, with nearest neighbor N1 atoms out of phase.

 $\omega = 424$ : These two modes have primarily in-phase N1
vibrations in the $x$-$y$ plane with Hf moving weakly against N1 and
  Ba and N2 in the Ba-N2 moving weakly in phase with N1. 

 $\omega = 468$ : N1 vibrating with large amplitude in phase with Ba, N2 and out of phase with Hf along the $z$ axis and each of the other atoms participate very weakly.

 $\omega = 623$ : These two modes are primarily out-of-phase N1
vibrations in the $x$-$y$ plane.

 $\omega = 641$ : This mode is primarily in-phase N2 vibrations
against other atoms along the $z$ axis. This mode dominates the ionic
response in the $z$ direction, especially due to the large magnitude
of BEC for N2: $Z^{*}_ {zz}$(N2) = -4.58.

 $\omega = 717$ : This mode is primarily out-of-phase N2 vibrations
along the $z$ axis with very weak participation from Ba and Hf.

\bibliographystyle{unsrt}
\bibliography{ref}

\end{document}